\documentclass[a4paper,11pt]{article}
\usepackage{pos}

\title{Searching for New Physics with $B^0_s \rightarrow D_s^{\pm} K^{\mp} $ Decays}
\author[a,b]{Robert Fleischer}
\author*[a]{Eleftheria Malami}

\affiliation[a]{Nikhef,\\
  Science Park 105, NL-1098 XG Amsterdam, Netherlands}

\affiliation[b]{Department of Physics and Astronomy, Vrije Universiteit Amsterdam,\\
NL-1081 HV Amsterdam, Netherlands}

\emailAdd{Robert.Fleischer@nikhef.nl}
\emailAdd{emalami@nikhef.nl}

\abstract{Particularly interesting processes to test the Standard Model are non-leptonic $B^0_s \rightarrow D_s^{\pm} K^{\mp}$ transitions. As these decays occur via pure tree diagrams, they allow a theoretically clean determination of the angle $\gamma$ of the unitarity triangle. Considering recent LHCb results, an intriguing picture arises, showing tension with the Standard Model. Utilising the available experimental data, we perform a theoretical analysis in order to shed more light on these puzzling patterns. Do these puzzles actually indicate footprints of New Physics?}

\FullConference{%
  *** Particles and Nuclei International Conference - PANIC2021 ***\\
  *** 5 - 10 September, 2021 ***\\
  *** Online ***
}

\begin{document}
\maketitle

\section{Standard Model Framework}
The $B^0_s\to D_s^\mp K^\pm$ decays are particularly interesting processes to test the Standard Model (SM) description of CP violation \cite{ADK,RF-BsDsK,DeBFKMST}. In the SM, these channels are generated only by tree diagram contributions. Due to $B^0_s$--$\bar B^0_s$ mixing, interference effects between the different decay paths arise, leading to a time-dependent CP asymmetry
\begin{equation}
 \frac{\Gamma(B^0_s(t)\to D_s^{+} K^-) - \Gamma(\bar{B}^0_s(t)\to D_s^{+} K^-) }{\Gamma(B^0_s(t)\to D_s^{+} K^-) + \Gamma(\bar{B}^0_s(t)\to D_s^{+} K^-) }   = \frac{{C}\,\cos(\Delta M_s\,t) + {S}\,\sin(\Delta M_s\,t)}
	{\cosh(y_s\,t/\tau_{B_s}) + {\cal A}_{\Delta\Gamma}\,\sinh(y_s\,t/\tau_{B_s})},
\end{equation}
with the observables $C$, $S$ and ${\cal A}_{\Delta\Gamma}$ and $y_s\equiv {\Delta\Gamma_s}/({2\,\Gamma_s})=0.062 \pm 0.004$. An analogous expression holds for the CP-conjugate final state $D_s^{-} K^+$, with the observables $\overline{C}$, $ \overline{S}$ and $\overline{{\cal A}}_{\Delta\Gamma}$.

Even though the $B^0_s\to D_s^\mp K^\pm$ channels are non-leptonic decays, thus challenging due to strong interactions, they allow a theoretically clean determination of the Unitarity Triangle (UT) angle $\gamma$. Inspired by an intriguing value of $\gamma$ reported by LHCb in \cite{BsDsK-LHCb-CP}, which is in tension with global SM analyses of the UT giving a value around $70^{\circ}$\cite{Amhis:2019ckw,PDG,CKMfitter,UTfit,LHCb:2021dcr}, we shed more light to this puzzling case \cite{Fleischer:2021cct, Fleischer:2021cwb}.  

Central role in our analysis plays the relation given by the product of two physical observables $\xi$ and $\bar\xi$ which measure the strength of the interference effects:
\begin{equation} \label{multxi}
 {\xi} \times \bar{\xi}= e^{-i2( \phi_s + \gamma)},
\end{equation} %
where $\phi_s$ is the $B^0_s$--$\bar{B}^0_s$ mixing phase. The parameters $\xi$ and $\bar\xi$ can be determined from the observables $C$, $S$, ${\cal A}_{\Delta\Gamma}$ and $\overline{C}$, $ \overline{S}$, $\overline{{\cal A}}_{\Delta\Gamma}$, respectively Consequently, Eq. \ref{multxi} allows the $ \phi_s + \gamma$ extraction and since $\phi_s$ can be determined through $B^0_s \rightarrow J/\psi \phi$ \cite{Barel:2020jvf}, we finally obtain $\gamma$. 

The LHCb collaboration performed a sophisticated fit to their data to extract $\gamma$ and other parameters, working under the assumption of $C+\bar{C}=0$, which holds in the SM. Using the current value $\phi_s=\left(-5^{+1.6}_{-1.5}\right)^\circ$ \cite{Barel:2020jvf}, which includes penguin corrections, we convert the LHCb result into
\begin{equation}\label{gamma-res-1}
\gamma=\left(131^{+17}_{-22}\right)^\circ .
\end{equation} 
In view of this surprisingly large value of $\gamma$, we need to transparently understand the situation. For this purpose, we utilise an expression for $\tan{(\phi_s + \gamma)}$ in terms of $S$, $ \overline{S}$ and ${\cal A}_{\Delta\Gamma}$, $\overline{{\cal A}}_{\Delta\Gamma}$ presented in \cite{Fleischer:2021cct, Fleischer:2021cwb}, allowing a simple extraction of $\phi_s + \gamma$. We find excellent agreement with the LHCb results, thereby confirming the intriguing picture. 

Could this puzzle arise from NP effects entering at the amplitude level of the  $B^0_s\to D_s^\mp K^\pm$ system? In order to answer this question, we look into the branching ratios. Due to the $B^0_s$--$\bar{B}^0_s$ oscillations, we need to distinguish between $\mathcal{B}_{\text{exp}}$ ``experimental" time-integrated branching ratios \cite{Dunietz:2000cr} and the $\mathcal{B}_{\text{th}} $  ``theoretical" ones where mixing effects are ``switched off" \cite{DeBFKMST,DeBruyn:2012wj}. In addition, we have to disentangle the interference effects between the two decay paths arising from the $B^0_s$--$\bar{B}^0_s$ mixing. We obtain the following individual theoretical branching ratios \cite{Fleischer:2021cct}:
\begin{align}\label{BRbar-Ds+K-}
\mathcal{B}(\bar B^0_s\to D_s^+K^-)_{\text{th}} &=2 \left[{|\xi|^2}/{\left(1+|\xi|^2 \right)} \right]\mathcal{B}_{\text{th}} = (1.94 \pm 0.21) \times 10^{-4}, \\
\mathcal{B}(B^0_s\to D_s^+K^-)_{\text{th}} &=2 \left[{1}/{\left(1+|\xi|^2 \right)} \right]\mathcal{B}_{\text{th}} =(0.26 \pm 0.12) \times 10^{-4}.
\end{align}

Factorization provides the theoretical framework to calculate the amplitudes and the branching ratios. A key example where 
``QCD factorization" is expected to work excellently is given by the color-allowed tree topology of the decay ${\bar{B}^0_s \rightarrow D_s^+ K^-}$\cite{Beneke:2000ry,bjor,DG,Neubert:1997uc,SCET}. The factorised amplitude is expressed in terms of Cabibbo--Kobayashi--Maskawa  (CKM) matrix elements, the kaon decay constant, the corresponding hadronic form factor and a parameter $a_{\rm 1 \, eff }$ which describes deviations from naive factorisation. Taking into account that the channels we study receive additional contributions from exchange topologies, the parameter $a_{\rm 1 \, eff }$ takes the form
\begin{equation}\label{a-eff-1-DsK}
a_{\rm 1 \, eff }^{D_s K}=a_{1}^{D_s K} \left(1+\frac{E_{D_s K}}{T_{D_s K}}\right) ,
\end{equation}
where $a_{1}^{D_s K}$ characterises non-factorisable effects entering the color-allowed tree amplitude $T_{D_s K}$, while $E_{D_s K}$ describes the non-factorisable exchange topologies. For our system \cite{Bordone:2020gao,Huber:2016xod},  we have the following values \cite{Fleischer:2021cct}:
\begin{equation}
|a_1^{D_sK}| = 1.07\pm0.02, \qquad |a_1^{K D_s}| = 1.1 \pm 0.1.
\label{eq:theo}
\end{equation}
Performing a detailed analysis, we find that no anomalous behaviour of the exchange topologies is indicated \cite{Fleischer:2021cct, Fleischer:2021cwb}, in line with other studies \cite{FST-BR}.
\begin{figure}[t!]
	\centering
	\includegraphics[width = 0.5\linewidth]{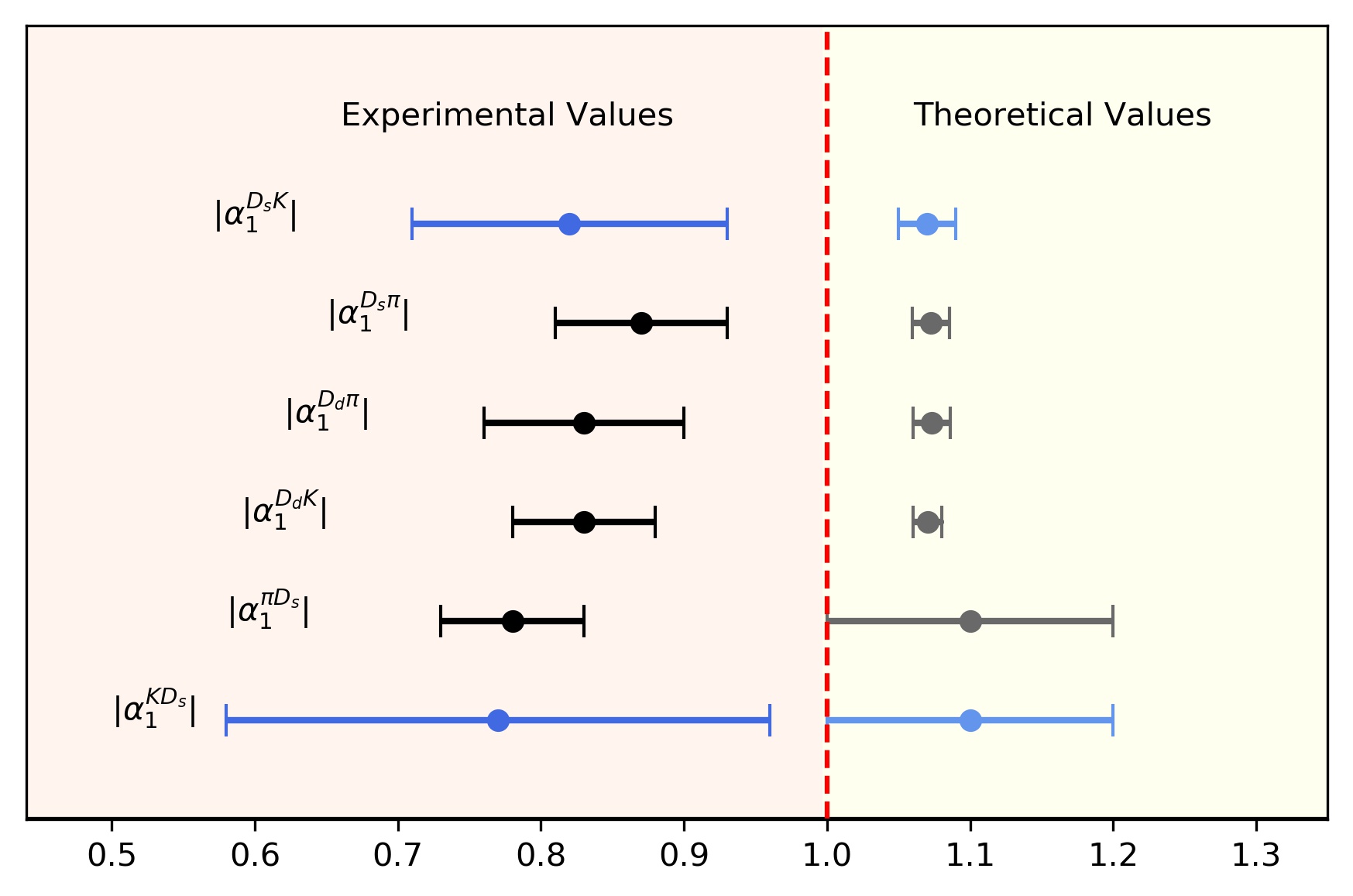}
	\caption{Experimental and theoretical SM values of the $|a_1|$ parameters for various decay processes.} \label{fig:aval}
\end{figure}

Our next step is to extract the $|a_1| $ parameters from the data in the cleanest possible way, where a key tool is provided by semileptonic decays \cite{Beneke:2000ry,FST-BR,Neubert:1997uc}.  Specifically for the $\bar{B}^0_s\to D_s^+K^-$ channel, we have the partner decay $\bar{B}^0_s \rightarrow D_s^{+}\ell^{-} \bar{\nu}_{\ell}$, and introduce ratios of the form
\begin{equation}
  R_{D_s^{+}K^{-}}\equiv\frac{\mathcal{B}(\bar{B}^0_s \rightarrow D_s^{+}K^{-})_{\rm th}}{{\mathrm{d}\mathcal{B}\left(\bar{B}^0_s \rightarrow D_s^{+}\ell^{-} \bar{\nu}_{\ell} \right)/{\mathrm{d}q^2}}|_{q^2=m_{K}^2}}=6 \pi^2 f_{K}^2 |V_{us}|^2 |a_{\rm 1 \, eff }^{D_s K}|^2  X_{D_s K}
    \label{semi}
\end{equation}
with $X_{D_s K}\equiv{\Phi_{\text{ph}}} [ {F_0^{B_s \rightarrow D_s}(m_K^2)}/{F_1^{B_s \rightarrow D_s}(m_K^2)}]^2$, where the phase-space factor ${\Phi_{\text{ph}}}$ is approximately 1. The advantage is that the $|V_{cb}|$ matix element cancels in the ratio and -- due to the normalization condition $F_0^{B_s \rightarrow D_s}(0)=F_1^{B_s \rightarrow D_s}(0)$ -- there is a minimal impact of the form factors. An analogous relation holds for the other decay channel. Finally, we obtain:
\begin{equation}
|a_{\rm 1}^{D_s K}| = 0.82 \pm 0.11, \qquad |a_{\rm 1}^{K D_s}| =0.77 \pm 0.19.
\label{eq:exp}
\end{equation}
Comparing these values with the theoretical predictions, we observe tension in the results, as the experimental values are surprisingly smaller. To complement the analysis, we had a detailed look at other $B_{(s)}$ decays with similar dynamics. Interestingly, we observe again a similar pattern.  In Fig.~\ref{fig:aval}, we illustrate the various $|a_1|$ values. We note that $\bar B^0_d\to D_d^+ K^-$ stands out, showing a discrepancy at the $4.8 \,\sigma$ level. This puzzling picture complements the intriguing $\gamma$ value following from the $B^0_s\to D_s^\mp K^\pm$ system. We also note that studies within physics beyond the SM have been performed in \cite{Iguro:2020ndk,Cai:2021mlt,Bordone:2021cca} and possible NP effects in non-leptonic tree-level $B$ decays were discussed in \cite{Brod:2014bfa,Lenz:2019lvd}.

\section{Towards New Physics}
In view of these puzzling results, we extend our analysis to include NP effects. We introduce the NP parameters
\begin{equation}
\bar{\rho} \, e^{i \bar{\delta}} e^{i \bar{\varphi}}  \equiv \frac{ A(\bar{B}^0_s \rightarrow D_s^+ K^-)_{{\text{NP}}} }{  A(\bar{B}^0_s \rightarrow D_s^+ K^-)_{{\text{SM}}} }, 
\end{equation}
where $\bar{\varphi}$ and $ \bar{\delta}$ denote CP-violating and CP-conserving phases, respectively. Similarly, we have $\rho$, $\varphi$ and $\delta$ for the CP-conjugate case. Generalising the SM relation $C=-\bar{C}$, assumed by LHCb, we arrive at
\begin{equation}
\xi \times \bar{\xi}  = \sqrt{1-2\left[\frac{C+\bar{C}}{\left(1+C\right)\left(1+\bar{C}\right)}
\right]}e^{-i\left[2 (\phi_s +\gamma_{\rm eff})\right]},
\end{equation}
where  the UT angle $\gamma$ enters as the ``effective" angle $\gamma_{\rm eff}\equiv \gamma+\frac{1}{2}\left(\Delta\Phi+\Delta\bar{\Phi}\right)= \gamma-\frac{1}{2}\left(\Delta\varphi+\Delta\bar{\varphi}\right)$.
Setting the strong phases $\delta$ and $\bar{\delta}$ to $0^\circ$, in agreement with the LHCb assumption, we apply our formalism to the current data. In the presence of NP contributions, we consider CP-averaged rates and introduce the quantities
\begin{equation}\label{b-bar-def}
\bar b \equiv \frac{\langle R_{D_s K} \rangle}{6 \pi^2 f_{K}^2 |V_{us}|^2 |a_{\rm 1 \, eff}^{D_s K}|^2 X_{D_s K} }
=1+2 \, \bar\rho\cos\bar\delta\cos\bar\varphi + \bar\rho^2 = 0.58 \pm 0.16
\end{equation}
\begin{equation}\label{b-def}
b\equiv \frac{\langle R_{K D_s}\rangle}{6 \pi^2 f_{D_s}^2 |V_{cs}|^2 |a_{\rm 1 \, eff}^{K D_s}|^2 X_{K D_s}}
= 1+2 \, \rho\cos\delta\cos\varphi + \rho^2 = 0.50 \pm 0.26,
\end{equation}
which in the SM are equal to 1. The deviations from this number reflect the puzzling patterns in Fig.~\ref{fig:aval}. Employing $\bar b$ and $b$, we can express the NP parameters in the following way:
\begin{equation} \label{extra}
        \bar{\rho} = - \cos{\bar{\varphi}} \pm \sqrt{\bar{b} -\sin^2\bar{\varphi}} , \qquad
      \rho =- \cos{\varphi} \pm \sqrt{b -\sin^2{\varphi}}.
\end{equation}

\begin{figure}[t!]
	\centering
\hspace{-0.8cm}\includegraphics[width = 0.315\linewidth]{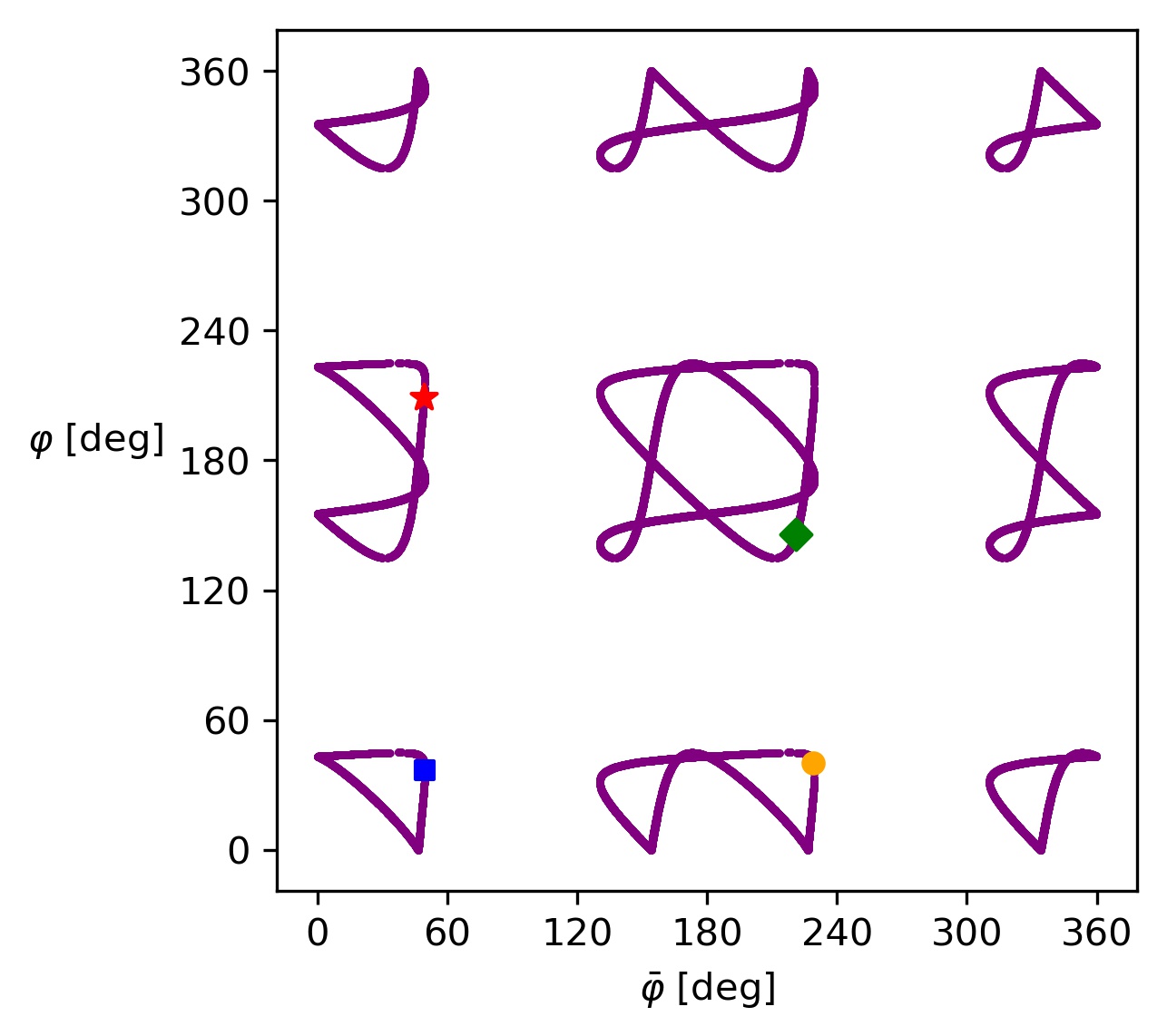}
\includegraphics[width = 0.285\linewidth]{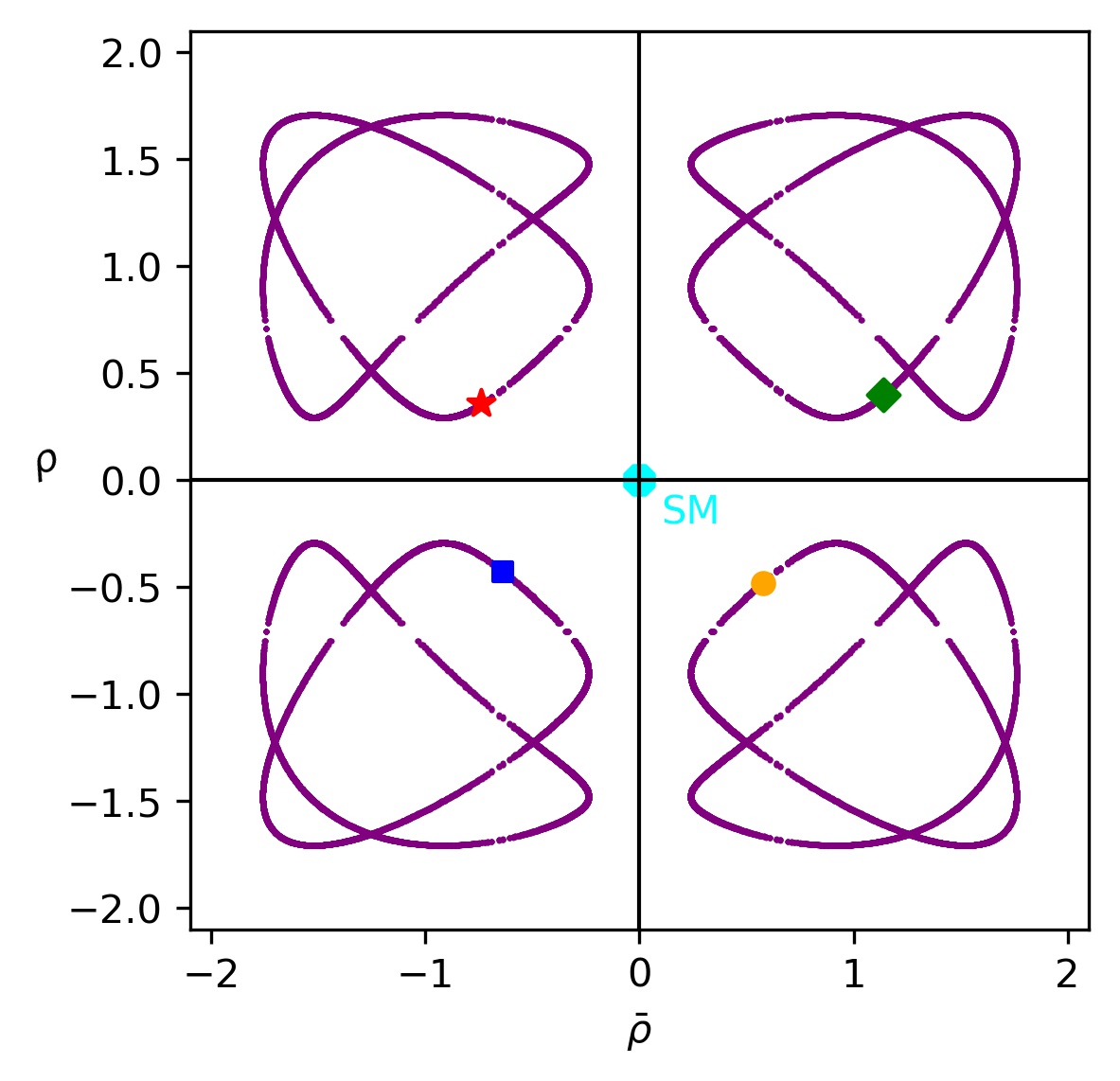}
\includegraphics[width = 0.34\linewidth]{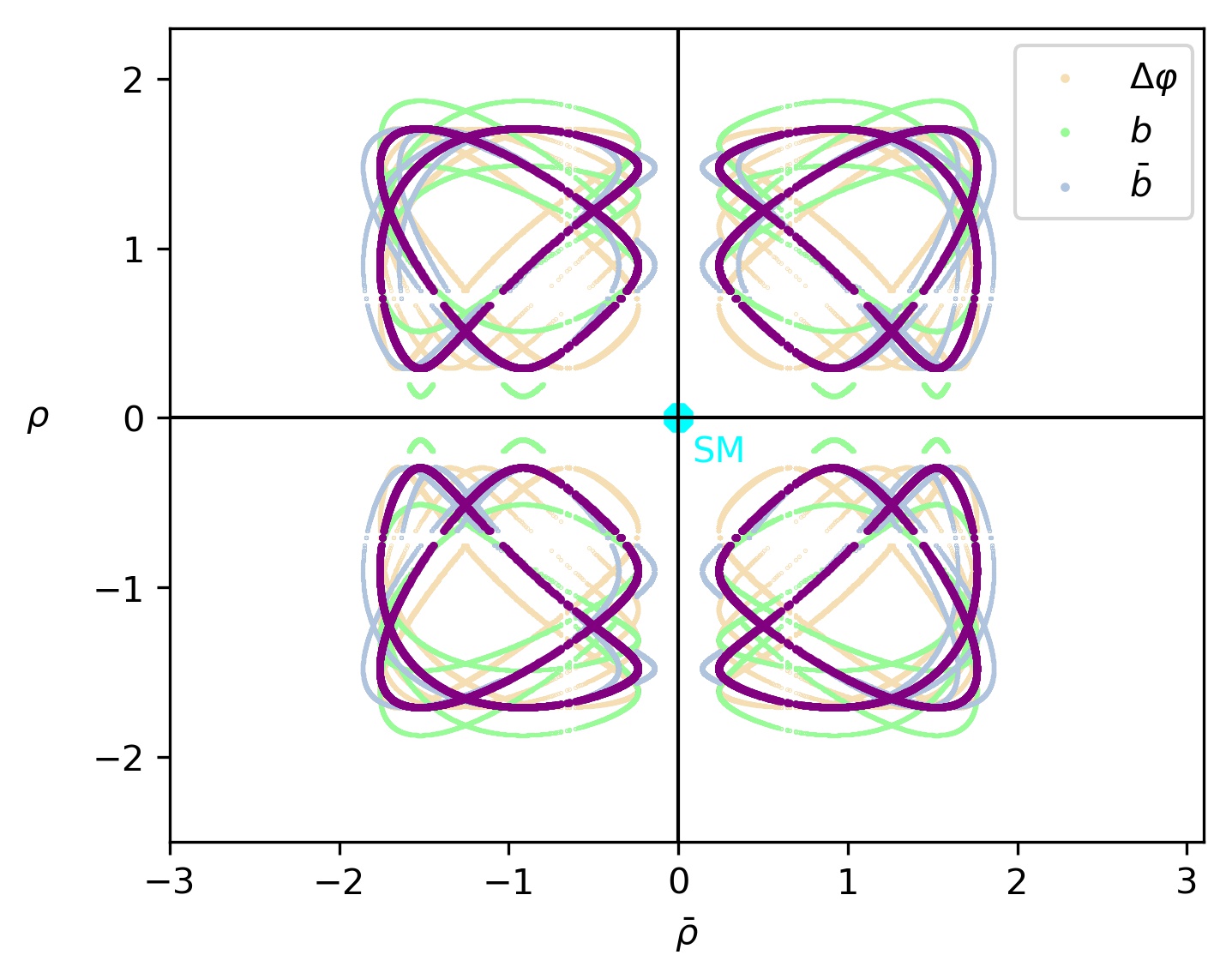}
\vspace*{-0.3truecm}
	\caption{Correlations for the central values of the current data in the $\bar{\varphi}$--$\varphi$ plane (left) and the $\bar{\rho}$--$\rho$ plane (center). As examples, we pick some points from the correlation in the $\bar\varphi$--$\varphi$ plane and show the corresponding values in the $\bar\rho$--$\rho$ plane (square, circle, diamond and star). Correlations in the $\bar{\rho}$--$\rho$ plane including uncertainties (left).}
	\label{fig:centrrhophi-1}
\end{figure}

Utilising the phase $\Delta\varphi=\Delta\bar{\varphi} = \gamma-\gamma_{\rm eff}=-(61 \pm 20)^\circ$ yields
\begin{equation}\label{gamma-res-10}
\tan\Delta\varphi=\frac{\rho\sin\varphi+\bar{\rho}\sin\bar{\varphi}+\bar{\rho}\rho\sin(\bar{\varphi}+\varphi)}{1+\rho\cos\varphi +
\bar{\rho}\cos\bar{\varphi}+\bar{\rho}\rho\cos(\bar{\varphi}+\varphi)} \ ,
\end{equation}
allowing us to obtain correlations between the NP parameters. Implementing Eq.~\ref{extra} in Eq.~\ref{gamma-res-10}, we may calculate ${\varphi}$ as a function of $\bar\varphi$, thereby fixing a contour in the ${\varphi}$--$\bar\varphi$ plane (right panel of Fig.~\ref{fig:centrrhophi-1}). Using again the expressions in Eq.~\ref{extra}, we may then also determine the correlation in the $\bar{\rho}$--$\rho$ plane (central plot in Fig.~\ref{fig:centrrhophi-1}), where each point is linked with $\bar\varphi$ and $\varphi$. Interestingly, we note that values as small as in the regime around 0.5 could accommodate the central values of the current data.

We also show the impact of the uncertainties of the input parameters $\Delta\varphi$, $b$ and $\bar{b}$, varying each one of them separately. The contours are denoted with lighter colours (left panel of Fig.~\ref{fig:centrrhophi-1}). We can now accommodate the data with NP contributions as small as about $30 \%$ of the SM amplitudes.

\section{Conclusions}
We have shown that in the $B^0_s\to D_s^\mp K^\pm$ system there are puzzling patterns both in CP violation, reflected by the value of $\gamma$, and in the branching ratios of the individual channels. We present a method to minimise the theoretical uncertainties, interpreting the branching ratio information in the cleanest possible way. Interestingly, the branching ratio results are consistent with patterns in decays with similar dynamics. 

In order to deal with this intriguing situation, we have developed a model-independent strategy to reveal NP effects. This strategy sets the stage for future analyses at the high-precision frontier. It will be exciting to see how the future data will evolve. This method may finally allow us to establish new sources of CP violation in the $B^0_s\to D_s^\mp K^\pm$ system.

\end{document}